\documentclass{amsart}\usepackage[]{graphicx}\usepackage[]{color}
\makeatletter
\def\maxwidth{ %
  \ifdim\Gin@nat@width>\linewidth
    \linewidth
  \else
    \Gin@nat@width
  \fi
}
\makeatother

\definecolor{fgcolor}{rgb}{0.345, 0.345, 0.345}
\newcommand{\hlnum}[1]{\textcolor[rgb]{0.686,0.059,0.569}{#1}}%
\newcommand{\hlstr}[1]{\textcolor[rgb]{0.192,0.494,0.8}{#1}}%
\newcommand{\hlcom}[1]{\textcolor[rgb]{0.678,0.584,0.686}{\textit{#1}}}%
\newcommand{\hlopt}[1]{\textcolor[rgb]{0,0,0}{#1}}%
\newcommand{\hlstd}[1]{\textcolor[rgb]{0.345,0.345,0.345}{#1}}%
\newcommand{\hlkwb}[1]{\textcolor[rgb]{0.69,0.353,0.396}{#1}}%
\newcommand{\hlkwc}[1]{\textcolor[rgb]{0.333,0.667,0.333}{#1}}%
\newcommand{\hlkwd}[1]{\textcolor[rgb]{0.737,0.353,0.396}{\textbf{#1}}}%

\usepackage{framed}
\makeatletter
\newenvironment{kframe}{%
 \def\at@end@of@kframe{}%
 \ifinner\ifhmode%
  \def\at@end@of@kframe{\end{minipage}}%
  \begin{minipage}{\columnwidth}%
 \fi\fi%
 \def\FrameCommand##1{\hskip\@totalleftmargin \hskip-\fboxsep
 \colorbox{shadecolor}{##1}\hskip-\fboxsep
     \hskip-\linewidth \hskip-\@totalleftmargin \hskip\columnwidth}%
 \MakeFramed {\advance\hsize-\width
   \@totalleftmargin\z@ \linewidth\hsize
   \@setminipage}}%
 {\par\unskip\endMakeFramed%
 \at@end@of@kframe}
\makeatother

\definecolor{shadecolor}{rgb}{.97, .97, .97}
\definecolor{messagecolor}{rgb}{0, 0, 0}
\definecolor{warningcolor}{rgb}{1, 0, 1}
\definecolor{errorcolor}{rgb}{1, 0, 0}
\newenvironment{knitrout}{}{} 

\usepackage{alltt}

\usepackage{amsmath,amssymb,amsthm}
\usepackage[utf8]{inputenc}
\usepackage{soul}
\usepackage{natbib}
\usepackage{geometry}
\usepackage[hidelinks]{hyperref} 
\usepackage{bbm} 
\newcommand{\Rpackage}[1]{{\texttt{#1}}}

\newcommand{\R}{{\normalfont\textsf{R }}{}}

\title{Valid Simultaneous Inference in High-Dimensional Settings (with the HDM Package for R)}
\thanks{Version:  \today}
\author{Philipp Bach, Victor Chernozhukov, Martin Spindler}
\IfFileExists{upquote.sty}{\usepackage{upquote}}{}
\begin{document}

\begin{abstract}
Due to the increasing availability of high-dimensional empirical applications in many research disciplines, valid simultaneous inference becomes more and more important. For instance, high-dimensional settings might arise  in economic studies due to very rich data sets with many potential covariates or in the analysis of treatment heterogeneities. Also the evaluation of potentially more complicated (non-linear) functional forms of the regression relationship leads to many potential variables for which simultaneous inferential statements might be of interest.
Here we provide a review of classical and modern methods for simultaneous inference in (high-dimensional) settings and illustrate their use by a case study using the R package \Rpackage{hdm}. 
The R package \Rpackage{hdm} implements valid joint powerful and efficient hypothesis tests for a potentially large number of coefficients as well as the construction of simultaneous confidence intervals and, therefore, provides useful methods to perform valid post-selection inference based on the LASSO.

\R and the package \Rpackage{hdm} are open-source software projects and can be freely downloaded from CRAN:
\texttt{http://cran.r-project.org}.

\end{abstract}

\maketitle

\pagestyle{myheadings}
\markboth{\sc Simultaneous Inference with HDM}{\sc }

\tableofcontents
\section{Introduction }
Valid simultaneous inference becomes very important when multiple hypotheses are tested at the same time. For instance, suppose a researcher wants to estimate a potentially large number of regression coefficients and to assess which of these coefficients are significantly different from zero at a given significance level $\alpha$. 
It is well known that in such a situation, the naive approach, i.e. simply ignoring multiple testing issues, will generally lead to flawed conclusions due to a large number of mistakenly rejected hypotheses. Indeed, the actual probability of incorrectly rejecting one or more hypotheses will in general exceed the claimed/desired level $\alpha$ by large. 

The statistical literature has proposed various approaches to mitigate the consequences of testing multiple hypotheses at the same time. These methods can be grouped into two approaches according to the underlying criterion. The first approach, initiated by the famous Bonferroni correction, seeks to control the probability of at least one false rejection which is called the \textit{family-wise error rate} (FWER). Since the definition of the FWER refers to the probability of making at least one type I error, the FWER-criterion is appealing from an intuitive point of view.
However, FWER control is often criticized to be conservative and, instead, the \textit{false discovery rate} (FDR) control is used as a criterion in many test procedures leading to the second major class of multiple testing correction methods, e.g. \citet{bh1995}. The FDR refers to the expected share of falsely rejected null hypotheses and, hence, results from FDR-procedures differ from classical tests results in terms of interpretation.  
Various approaches aim at maintaining control of the FWER and reducing conservativeness at the same time by incorporating a stepwise procedure, for instance, the stepdown method of \cite{holm1979}. Moreover, taking the dependence structure of test statistics into consideration allows to reduce conservativeness of FWER-procedures, as in the stepdown procedure of  \cite{romanowolf2005, romanowolf2} which is based on resampling methods.

In the following, we review classical methods and will describe how valid simultaneous inference can be conducted in a high-dimensional regression setting, i.e. if the number of covariates exceeds the number of observations, and give examples how the presented methods can be applied with  the statistical software \R{.}  It is well-known that classical regression methods, such as ordinary least squares, break down in high-dimensional settings. Instead, regularization methods, e.g. the lasso, can be used for estimation. However, post-selection inference is non-trivial and requires modification of the estimators. 

The \R  package \Rpackage{hdm} implements the double-selection approach of \cite{BCH2011restud} that allows to perform valid inference based on model-selection with the lasso. Moreover,  \Rpackage{hdm} provides powerful tests for a large number of hypotheses using the multiplier bootstrap, potentially in combination with the Romano and Wolf stepdown procedure (starting with Version \texttt{0.3.0}) . The implemented methods are less conservative / more powerful than traditional approaches as they allow to take the dependence structure among test statistics into account and proceed in a stepwise manner. At the same time, the implemented procedures are attractive from an intuitive point of view as they guarantee control of the \textit{family-wise error rate} (FWER). 

The remainder of the paper is organized as follows. First, the setting is introduced and an overview on valid post-selection inference in high dimensions is provided. Second, a short and selective review on traditional and recent methods to adjust for multiple testing is presented. Third, we give an overview on the tools for valid post-selection inference in high-dimensions available in the \R package \Rpackage{hdm}. Fourth, the use of the software is illustrated in a simulated and a replicable real-data example. A conclusion is provided in the last section.

\section{The Setting}

We are interested in testing a set of $K$ hypotheses $H_1, ..., H_K$ in a high-dimensional regression model, i.e. a regression where the number of covariates $p$ is large, potentially much larger than the number of observations $n$, i.e. we have $p\gg n$. The ultimate objective in this setting is to perform inference on a subset of the regression coefficients, i.e. a vector of so-called ``target'' coefficients $\theta_k$ with $k = 1,..., K$ and $K<n$.\footnote{The setting with a potentially infinite-dimensional vector of target coefficients is theoretically considered in \cite{belloni2017} and \cite{BCK2015}.} 
 \begin{align} \label{model}
 y_i &= \beta_0 + d_i'\theta +  x_i' \beta + \epsilon_i, \quad \quad i = 1,\ldots, n,
 \end{align}
where $\beta_0$ is an intercept and $\beta$ denote the regression coefficients of the control variables $x_i$. Moreover, it is assumed that $E_n[\epsilon_i x_i] = 0$. In this setting, $K$  hypotheses are tested for the coefficients that correspond to the effect of the ``target'' variables $d_i$ on the outcome $y_i$
\begin{align} \label{hyp}
H_{0,k} : \theta_k = 0, \quad \quad  k=1,..., K .  
\end{align}
For instance, such a high-dimensional regression setting arises in causal program evaluation studies, where a large number of regressors is included to approximate a potentially complicated,  non-linear  population regression function using transformations with dictionaries, e.g. interactions, splines or polynomials. Alternatively, an analysis of heterogeneous treatment effects across possibly many subgroups as in the example in Section \ref{example} might require a large number of interactions of the regressors. 

Suppose, there is a test procedure for each of the hypotheses leading to test statistics $t_1, ..., t_K$ and unadjusted p-values $p_1, ..., p_K$. In the context of multiple testing, it is often helpful to sort the p-values in an increasing order (i.e. the ``most significant'' test result as the first in the row) and the hypotheses likewise, i.e. $p_{(1)}, ..., p_{(K)}$ and $H_{(1)}, ..., H_{(K)}$ with $p_{(1)} \le p_{(2)} \le ... \le p_{(K)}$. Also the test statistics are ordered by the same logic $\vert t_{(1)} \vert \ge \vert t_{(2)} \vert \ge ... \ge \vert t_{(K)} \vert$. A researcher decides whether to accept or to reject a null hypothesis if the corresponding p-value $p_k$ is above or below a prespecified significance level $\alpha$. Generally, the significance level corresponds to the probability of erroneously rejecting a true null hypothesis. 
However, if the conclusions are based on a comparison of unadjusted p-values and the significance level, the probability of incorrectly rejecting at least one of the hypotheses will, potentially by large, exceed the claimed level $\alpha$. Hence, adjustment for multiple testing becomes necessary to draw appropriate inferential conclusions.

\section{Simultaneous Post-Selection Inference in High Dimensions -- An Overview}

In high-dimensional settings traditional regression methods such as ordinary least squares break down and testing the $K$ hypotheses will severely suffer from the shortcomings of the underlying estimation method. Penalization methods, for instance the lasso or other machine learning techniques, provide an opportunity to overcome the failure of traditional least squares estimation as they regularize the regression problem in Equation \ref{model} by introducing a penalization term. In the example of lasso, the ordinary least squares minimization problem is extended by a penalization of the regression coefficients using the $l_1$-norm. The lasso estimator is the solution to the maximization problem 
\begin{align}
\left(\hat{\theta}, \hat{\beta}\right) = \arg \min_{\theta, \beta} 	\mathbb{E}_n \left[ \left(y_i-\beta_0 - d_i'\theta - x_i'\beta \right)^2  \right] + \frac{\lambda}{n} \left\|\hat{\psi} \left( \theta', \beta' \right)' \right\|_1,
\end{align} 
with $\| \bullet \|_1$ being the $l_1$-norm, $\lambda$ is a penalization parameter and $\hat{\psi}$ denotes a diagonal matrix of penalty loadings. More details on the choice of $\lambda$ and $\hat{\psi}$ as implemented in the \Rpackage{hdm} package can be found in the package vignette available at \href{https://cran.r-project.org/web/packages/hdm/index.html}{CRAN} and \cite{rjournalhdm}.
As a consequence of the $l_1$-penalization, some of the coefficients are shrunk towards zero and some of them are set exactly equal to zero. In general, inference after such a selection step is only valid under the strong assumption of perfect model selection, i.e. the lasso does only set those coefficients to zero that truly have no effect on $y_i$. However perfect model selection and the underlying assumptions are often considered unrealistic in real-world applications leading to a breakdown of the \textit{naive} inferential framework and, thus, flawed inferential conclusions.

In contrast to the naive procedure, the so-called double-selection approach of \cite{BCH2011restud} tolerates small model selection mistakes so that valid confidence intervals and test procedures can be based on  the lasso. 
The double-selection method is based on orthogonalized moment equations and introduces an auxiliary (lasso) regression step for each of the target coefficients in order to avoid that variable selection erraneously excludes variables that are related to both the outcome and the target regressors by setting their coefficients equal to zero. 
 Double selection proceeds as follows: (1) For each of the target variables in $d_{j,i}$, $j = 1,\ldots,K$, a lasso regression is estimated to identify the most important predictors among the covariates $x_i$ and the  remaining target variables $d_{-j,i}$. (2) A lasso regression of the outcome variable $y_i$ on all explanatory variables, except for $d_{j,i}$, is estimated to identify predictors of $y_i$. This step is executed for each of the target variables $d_j$ with $j = 1,\ldots, K$. (3) The target coefficients $\theta$ are estimated from a linear regression of the outcome on all target variables as well as all covariates that have been selected in either step (1) or (2). 
As a consequence of the double-selection procedure, the risk of an omitted variable bias that might arise due to imperfect model selection is reduced. 
It can be shown that the double-selection estimator $\hat{\theta}^{DS}_k$ is asymptotically normally distributed under a set of regularity assumptions. Probably, the most important of these assumptions is (approximate) sparsity. This assumption states that only a subset of the regressors suffice to describe the relationship of the outcome variable and the covariates and that all other regressors have no or only a negligible effect on the outcome. 
In general, valid post-selection inference is compatible with other tools from the machine learning literature, for instance elastic nets or regression trees, as long as these methods satisfy some regularity conditions \citep{BCK2015, BCH2011restud}.

In the multiple testing scenario described above, the double-selection approach can be used to test the $K$ null hypotheses $H_{0,k}$, $k=1,...,K$. \citet{BCK2015} show that a valid ($1-\alpha$)-confidence interval can be constructed by using the multiplier boostrap as established in \citet{CCK2013}. Moreover, \citet{CCK2013} and  \citet{BCK2015} show that a multiplier bootstrap version of the Romano-Wolf method can be used to construct a joint significance test in a high-dimensional setting such that asymptotic control of the FWER is obtained.

An advantage of the stepdown method of \cite{romanowolf2005, romanowolf2} is that it is based on resampling methods. Hence, it is able to account for the dependence structure underlying the test statistics and to give less conservative results as compared to methods such as the Bonferroni and Holm correction. The  idea of the Romano-Wolf procedure is to construct rectangular simultaneous confidence intervals in subsequent steps whereas in each step, the coverage probability is kept above a level of $(1-\alpha)$. If in step $j$, the confidence set does not contain zero in dimension $k$, the corresponding $H_k$ is rejected. In step $j+1$, the algorithm proceeds analogously by constructing a rectangular joint confidence region for those coefficients for which the null hypotheses has not been rejected in step $j$ or before. The algorithm stops if no hypothesis is rejected anymore. To take the dependence structure of the test hypotheses into account, the classical Romano-Wolf stepdown procedure uses the bootstrap to compute the constant $c_{(1-\alpha)}$ which is needed to construct a rectangular confidence interval. This constant is estimated by the ($1-\alpha$)-quantile of the maxima of the bootstrapped test statistics in each step to guarantee the coverage probability of ($1-\alpha$).
 The computational burden of the Romano-Wolf stepdown procedure can be reduced by using the multipler bootstrap. This bootstrap method requires the calculation of the solution of an orthogonal moment equation only once and then operates on pertubations by realizations of an independently and standard normally distributed random variable \citep{CCK2013}.  %

\section{Methods for Testing multiple Hypotheses}  \label{overview}

\subsection{A Global Test for Joint Significance with Lasso Regression}

A basic question frequently arising in empirical work is whether the Lasso regression has explanatory power, comparable to a F-test for the classical linear regression model. The construction of a joint significance test follows \citet{CCK2013} (Appendix M). Based on the model $ y_i = \beta_0 + d_i'\theta + x_i'\beta + \epsilon_i $ with intercept $\beta_0$, the null hypothesis of joint statistical in-significance is  $H_0: (\theta', \beta')'=\mathbf{0}$. 
The null hypothesis implies that
 $$ \mathbb{E} \left[ (y_i - \beta_0) x_i \right] = 0,$$
 and the restriction can be tested using the sup-score statistic:
 $$S = \| \sqrt{n} \mathbb{E}_n \left[ (y_i - \hat \beta_0) x_i \right] \|_\infty,$$
where $\hat \beta_0 =  \mathbb{E}_n [y_i]$.  The critical value for this statistic can be approximated  by the multiplier bootstrap procedure, which simulates the statistic:
 $$ S^* = \| \sqrt{n} \mathbb{E}_n \left[ (y_i - \hat \beta_0) x_i g_i \right] \|_\infty,$$
 
where $g_i$'s are i.i.d. $N(0,1)$, conditional on the data. The $(1-\alpha)$-quantile of $S^*$ serves as the critical value, $c(1-\alpha)$. We reject the null if $S > c(1-\alpha)$ in favor of statistical significance, and we keep the null of non-significance otherwise.

\subsection{Multiple Hypotheses Testing with Control of the Familywise Error Rate}
The FWER is defined as the probability of falsely rejecting at least one hypothesis. The goal is to control the FWER and to secure that it does not exceed a prespecified level $\alpha$. We assume that for the individual tests the significance level is set uniformly to $\alpha$.  

\subsubsection{Bonferroni Correction}
According to the Bonferroni correction the cutoff of the p-values is set to $\alpha^{*}=\alpha/K$ and all hypotheses with p-values below the adjusted level $\alpha^{*}$ are rejected. Bool's inequality then gives directly that the FWER is smaller or equal to $\alpha$. Instead of adjusting the level of $\alpha$ to $\alpha^{*}$, it is possible to adjust the p-values so that we reject a hypothesis $H_k$ if $p^{*}_k = K \cdot p_{k} < \alpha$. 
A drawback of the procedure is that it is quite conservative, meaning that in many applications, in particular in high-dimensional settings when many hypotheses are tested simultaneously, often no or very few hypotheses are rejected, increasing the risk of accepting false null hypotheses (i.e. of a type II error). 

\subsubsection{Bonferroni-Holm Correction}
We again assume that the p-values are ordered (from lowest to highest) $p_{(1)} \le \ldots \le p_{(K)}$ with corresponding hypotheses $H_{(1)}, \ldots, H_{(K)}$. The Bonferroni-Holm procedure controls the FWER by the following procedure: Let $k$ be the smallest index such that the corresponding p-value exceeds the adjusted cutoff $\alpha^{*}$. 
\begin{align*}
k = \min_{j} \{ p_{(j)} > \underbrace{\frac{\alpha}{K- j + 1}}_{\alpha^{*}} \},
\end{align*}

Reject the null hypothesis $H_{(1)}, \ldots, H_{(k-1)}$ and accept $H_{(k)}, \ldots, H_{(K)}$. The Bonferroni-Holm procedure can be considered as \textit{general improvement} over the Bonferroni correction that maintains control of the FWER and reduces the risk of a type II error at the same time. The adjusted p-value according to the Bonferroni-Holm correction are computed as $p_{(j)}^{*} = \max_{l \le j} \min \{(K-j+1)p_{(j)}, 1 \}$ with $l = 1, \ldots, j$.

\subsubsection{Joint Confidence Region Using Multiplier Bootstrap}
 \citet{BCK2015} derive valid ($1-\alpha$)-confidence regions for the vector of target coefficients, $\theta$,  in the high-dimensional regression setting in Equation \ref{model} estimated with lasso. 
  The confidence regions which are constructed with the multiplier bootstrap can be used equivalently to a joint signficance test of the $K$ hypotheses. Accordingly, the null hypotheses $H_{0,k}: \theta_{k} = 0$, $k = 1, \ldots, K$, would be rejected at the level $\alpha$ if the simultaneous ($1-\alpha$)-confidence region does not cover zero in dimension $k$.

\subsubsection{Romano-Wolf Stepdown Procedure}
While the Bonferroni and Bonferroni-Holm correction do not take into account the dependence structure of the test statistics, further improvements in the control of the type II error can be achieved by modeling the dependence structure using resampling. A very popular and powerful method in this regard is the Romano-Wolf stepdown procedure. Stepdown methods proceed in several rounds where in each round a decision is taken on the set of hypotheses being rejected. The  algorithm continues until no further hypotheses are rejected. 
The Romano-Wolf stepdown procedure guarantees asymptotic control of the FWER at level $\alpha$ by constructing a sequence of simultaneous tests.
We present the recent version of the Romano-Wolf method from \citet{CCK2013} and  \citet{BCK2015} who prove the validity of the procedure in combination with the multiplier bootstrap.


\vspace{0.5cm} 

 \begin{enumerate} 
 
 \item[1)] Sort the test statistics in a decreasing order (in terms of their absolute values):
 \begin{align*}
 | t_{(1)} | \ge | t_{(2)} | \ge .... \ge | t_{(K)}|.
 \end{align*}
 \item[2)] Draw $B$ multiplier bootstrap versions for each of the test statistics $t_{(k)}^{*,b}$, $b=1, ..., B$, and $k=1,..., K$,
 \item[3)] For each $b$ and $k$ determine the maximum of the bootstrapped test statistics $m(t_{(k)}^{*,b}) = \max \left\{|t_{(k)}^{*,b}|, |t_{(k+1)}^{*,b}|, ..., |t_{(K)}^{*,b}| \right\}$.  

 \item[4)] Compute initial p-values, for $(k)=1, ..., K$
  \begin{align*}
\hat{p}^{init}_{(k)} := \frac{ \sum_{b=1}^{B} \mathbbm{1}  \{ m(t_{(k)}^{*,b}) \ge | t_{(k)}| \}}{B}
 \end{align*}
  
  \item[5)] Compute adjusted p-values by ensuring monotonicity
    \begin{itemize}
 \item[a)] if $(k)=1$
 \begin{align*}
 \hat{p}^{*}_{(1)}:= \hat p^{init}_{(1)}
 \end{align*}
 \item[b)] if $(k)=2, ..., K$
 \begin{align*}
 \hat{p}^{*}_{(k)} := \max\{\hat{p}^{init}_{(k)}, \hat{p}^{*}_{(k-1)}\}
 \end{align*}
 \end{itemize}
 
 \end{enumerate}
 
%

The p-adjustment algorithm parallels that in \cite{romanowolf2016} with the only difference that the bootstrap test statistics are computed efficiently with the multiplier bootstrap procedure instead of the classical bootstrap. In contrast to traditional bootstrap methods, the multiplier bootstrap does not require re-estimation of the lasso regression for each bootstrap sample. \cite{romanowolf2, romanowolf2016} recommend a high number of bootstrap repetitions $B\ge 1000$. 

If the data stem from a randomized experiment, the method introduced in \cite{list2016} can be used. It is a variant of the Romano-Wolf procedure under uncounfoundedness. Moreover, it allows to compare the effect of different treatments and several outcome variables simultaneously.


\subsection{Multiple Hypotheses Testing with Control of the False Discovery Rate}
The FWER is a very strict criterion which is often very conservative. This means that in settings when thousands or hundred thousands of hypotheses are tested simultaneously, the FWER does often not detect useful signals. Hence, in large-scale settings frequently a less strict criterion, the so-called false discovery rate (FDR) is employed. The false discovery proportion (FDP) is defined as the ratio of the number of hypotheses which are wrongly classified as significant (false positives) and the total number of positives. If the latter is zero, it is defined as zero. The FDR is defined as the expected value of the FDP : $FDR=\mathbb{E}(FDP)$. The FDR concept reflects the tradeoff between false discoveries and true discoveries.

\subsubsection{Benjamini-Hochberg Procedure}
To control the FDR, the Benjamini-Hochberg (BH) procedure ranks the hypotheses according to the corresponding p-values and then chooses a cutoff along the ranking to control the FDR at a prespecified level of $\gamma \in (0,1)$. The BH procedure first uses a stepup comparision to find a cutoff p-value:
%
%
\begin{align*}
k = \max_{j} \{ p_{(j)} \leq j \frac{\gamma}{K} \},
\end{align*}
and then rejects all hypotheses $H_{j}, j=1,\ldots,k$. 
In most applications, $\gamma = 0.1$ is chosen. 

\section{Implementation in R}

Estimation of the high-dimensional regression model in Equation \ref{model} and simultaneous inference on the target coefficients is implemented in the R package \Rpackage{hdm}  available at \href{https://cran.r-project.org/web/packages/hdm/index.html}{CRAN}. 
\Rpackage{hdm} provides an implementation of the double-selection approach of \citet{BCK2015} using the lasso as the variable selection device. The function \texttt{rlassoEffects()} does valid inference on a specified set of target parameters and returns an object of S3 class \texttt{rlassoEffects}. This output object is used to perform \textit{simultaneous} inference subsequently as described in the following. 
 More details on the \Rpackage{hdm} package and introductory examples are provided in the \Rpackage{hdm} vignette available at \href{https://cran.r-project.org/web/packages/hdm/index.html}{CRAN} and \cite{rjournalhdm}.
The package \Rpackage{hdm} offers three ways to perform valid simultaneous inference in high-dimensional settings: 

\begin{enumerate}

\item \textbf{ Overall Significance Test } 

\noindent
The \Rpackage{hdm} provides a joint significance test that is comparable to a F-test known from classical ordinary least squares regression. Based on \citet{CCK2013} (Appendix M), the null hypothesis that no covariate has explanatory power for the outcome $y_i$ is tested, i.e.
\begin{align*}
H_0: (\theta', \beta')' = \mathbf{0}.
\end{align*}
The test is performed automatically if \texttt{summary()} is called for an object of the S3 class \texttt{rlasso}. This object corresponds to the output of the function \texttt{rlasso()} which implements the lasso estimator using a theory-based rule for determining the penalization parameter. 
\\  
\item \textbf{ Joint Confidence Interval }

\noindent
 Based on an object of the S3 class \texttt{rlassoEffects}, a valid joint confidence interval with coverage probability $(1-\alpha)$ can be constructed for the specified target coefficients using the command \texttt{confint()} with the option \texttt{joint=TRUE}. 
\\
\item \textbf{Multiple Testing Adjustment of p-Values}

\noindent
Starting with Version \texttt{0.3.0}, the \Rpackage{hdm} package offers the S3 method \texttt{p\_adjust()} for objects inheriting from classes \texttt{rlassoEffects} and \texttt{lm}.  By default, \texttt{p\_adjust()} implements the Romano-Wolf stepdown procedure using the computationally efficient multiplier bootstrap procedure (option \texttt{method = "RW"}). Hence, the \Rpackage{hdm} offers an implementation of the p-value adjustment that corresponds to a joint test \`{a} la Romano-Wolf for both   post-selection inference based on double selection with the lasso as well as for ordinary least squares regression. Moreover, the \texttt{p\_adjust()} call offers classical adjustment methods in a user-friendly way, i.e. the function can be executed directly on the  output object returned from a regression with \texttt{rlassoEffects()} or \texttt{lm()}. The hosted correction methods are the methods provided in the \texttt{p.adjust()} command of the basic \Rpackage{stats} package, i.e. Bonferroni, Bonferroni-Holm, and Benjamini-Hochberg among others.  If an object of class \texttt{lm} is used, the user can provide an index of the coefficients to be tested simultaneously. By default, all coefficients are tested. 

\end{enumerate}

\section{A Simulation Study for Valid Simultaneous Inference in High Dimensions}

The simulation study provides a finite-sample comparison of different multiple testing corrections in a high-dimensional setting, i.e. the Bonferroni method, the Bonferroni-Holm procedure, Benjamini-Hochberg adjustment and the Romano-Wolf stepdown method. In addition, the study illustrates the failure of the \textit{naive} approach that ignores the problem of simultaneous hypotheses testing, i.e. without any correction of the significance level or p-values. 

\subsection{Simulation Setting}
We consider a regression of a continuous outcome variable $y_i$ on a set of $K = 60$ regressors, $d_i$,
\begin{align} \label{regsim}
y_i = \beta_0 + d_i'\theta + \epsilon_i , \quad \quad i = 1,\ldots, n,
\end{align}
with $\varepsilon_i \sim N(0,\sigma^2)$ and variance $\sigma^2 =  3$. In our setting, the realizations of $d_i$ are generated by a joint normal distribution $d_i \sim N(\mu, \Sigma)$ with $\mu = \mathbf{0}$ and $\Sigma_{j,k} = \rho^{|j-k|}$ with $\rho = 0.9$. We consider the case of an i.i.d. sample with $n = 200$ and $n = 500$ observations. The setting is sparse in that only few, $s = 12$, regressors are truly non-zero whereas the location of the non-zero coefficient is only known by an inaccessible oracle. Thus, the lasso is used to select the set of explanatory variables with a non-zero coefficient. Figure \ref{betaplot} presents the regression coefficients in the simulation study. 

\begin{knitrout}\footnotesize
\definecolor{shadecolor}{rgb}{1, 1, 1}\color{fgcolor}\begin{figure}

{\centering \includegraphics[width=\linewidth]{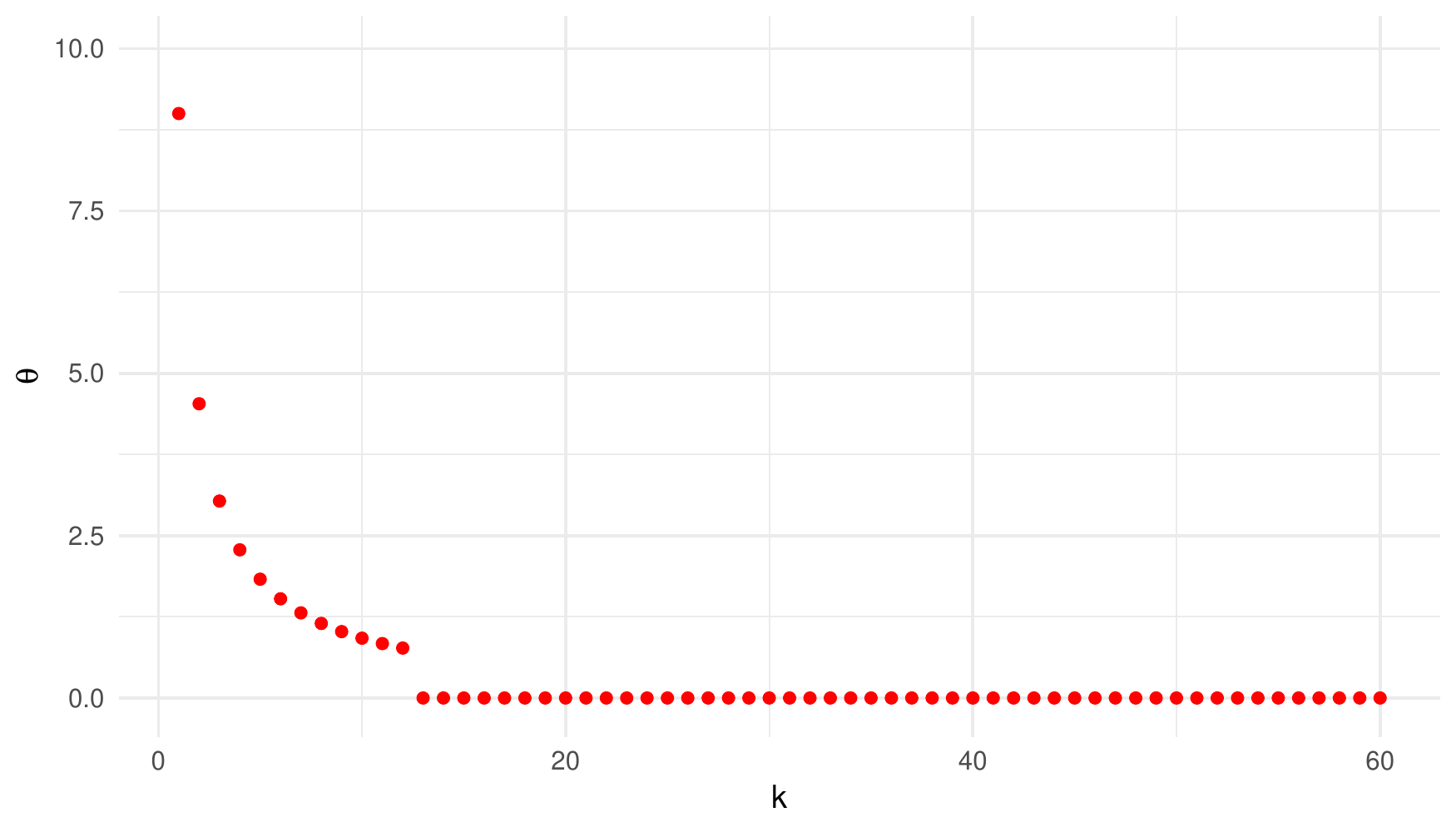} 

}

\caption[\label{betaplot} Regression Coefficients, Simulation Study]{\label{betaplot} Regression Coefficients, Simulation Study.}\label{fig:unnamed-chunk-2}
\end{figure}

\end{knitrout}
The regression in Equation \ref{regsim} is estimated wih post-lasso and inference for all regression coefficients is based on double selection.\footnote{More details on implementation of the simulation study are provided in the appendix and the supplemental material available at \url{https://www.bwl.uni-hamburg.de/en/statistik/forschung/software-und-daten.html}.} The $K=60$ hypotheses are tested simultaneously
\begin{align*} 
H_{0,k} : \theta_k = 0, \quad \quad  k=1,..., K .  
\end{align*}

\subsection{Results}
The simulation results are summarized in Table \ref{simresults1}. The reported results refer to averages over $R=5000$ repetitions in terms of correct and incorrect rejections of null hypotheses at a prespecified level of $\alpha = 0.1$  as well as the empirical FWER and FDR.

The results show that multiple testing adjustment is of great importance since naive inferential statements might be invalid. If each of the hypotheses is tested at a  significance level of $\alpha = 0.1$ without adjustment of the p-values, the naive procedure leads to at least one incorrect rejection with a probability of almost 1. Also the Benjamini-Hochberg procedure with $\gamma = 0.1$  incorrectly rejects a true null in more than 3000 out of the 5000 repetitions. On average, more than 5 ($n=200$) and more than 4 ($n=500$) true hypotheses are rejected without adjustment of p-values.  The simulation study illustrates that control of the FDR is achieved by the Benjamini-Hochberg correction. Accordingly, one would incorrectly reject more than 1 hypothesis on average. Over all 5000 repetitions 9.5\% ($n=200$) and 8.7\% ($n=500$) of all rejections are incorrect (false positives) which is below the specified level of $\gamma = 0.1$.

Methods with asymptotic control of the FWER are much less likely to erraneously reject true null hypotheses. In the setting with $n=200$ observations, the number of incorrect rejections is on average around 0.1 to 0.17 with the Bonferroni correction being most conservative. The empirical familywise error rates are very close to the desired level $0.1$, despite the small number of observations and the relatively large number of tested hypotheses. With larger sample size ($n=500$) the empirical FWERs approach the level $0.1$. The price for control of the probability of at least one type I error is paid in terms of power with less correct rejections for the FWER-methods as compared to the Benjamini-Hochberg correction. The largest number of correct rejections while still maintaining control of the FWER is achieved by the Romano-Wolf stepdown procedure. Hence, taking the dependence structure of the test statistics into account is favorable in case of dependencies among test statistics. 



\begin{table}[t]
\begin{tabular}{l r r r r r } \hline \hline 
   & Naive & Benjamini-Hochberg & Bonferroni & Bonferroni-Holm & Romano-Wolf \\[0.8ex]
     \hline   \hline 
   \multicolumn{6}{c}{Correct Rejections} \\[0.8ex] \hline 
 $n = 200$ & $10.749$ & $9.375$ & $7.597$ & $7.674$ & $7.765$ \\
  & $(0.798)$ & $(0.976)$ & $(0.965)$ & $(0.966)$ & $(0.963)$  \\
  & & & & & \\
  \hline 
\multicolumn{6}{c}{Incorrect Rejections}  \\ \hline 
    & $5.114$ & $1.128$ & $0.129$ & $0.145$ & $0.163$ \\
  & $(2.386)$ & $(1.271)$ & $(0.372)$ & $(0.397)$ & $(0.425)$ \\
  & & & & & \\
  \hline 
  \multicolumn{6}{c}{Familywise Error Rate}  \\ \hline 
     & $0.991$ & $0.606$ & $0.117$ & $0.129$ & $0.143$ \\
 & & & & & \\
  \hline 
  \multicolumn{6}{c}{False Discovery Rate}  \\ \hline 
     & $0.308$ & $0.095$ & $0.015$ & $0.016$ & $0.018$ \\
   & & & & & \\ \hline   \hline
  \multicolumn{6}{c}{Correct Rejections} \\[0.8ex] \hline
 $n = 500$ & $11.872$ & $11.578$ & $10.658$ & $10.735$ & $10.785$ \\
  & $(0.338)$ & $(0.558)$ & $(0.745)$ & $(0.742)$ & $(0.738)$  \\
  & & & & & \\
  \hline
\multicolumn{6}{c}{Incorrect Rejections}  \\ \hline
    & $4.893$ & $1.226$ & $0.101$ & $0.121$ & $0.137$ \\
  & $(2.381)$ & $(1.317)$ & $(0.333)$ & $(0.362)$ & $(0.387)$ \\
  & & & & & \\ 
  \hline 
  \multicolumn{6}{c}{Familywise Error Rate}  \\ \hline 
     & $0.987$ & $0.636$ & $0.091$ & $0.110$ & $0.123$ \\
    & & & & & \\
  \hline 
  \multicolumn{6}{c}{False Discovery Rate}  \\ \hline 
    & $0.278$ & $0.087$ & $0.009$ & $0.010$ & $0.011$  \\[0.8ex] \hline   \hline
\end{tabular}
\caption{Simulation Results}
\label{simresults1}
The Table presents the average number of correct or incorrect rejections at a significance level $\alpha = 0.1$ and the FWER over all $R = 5000$ repetitions. For the  Benjamini-Hochberg adjustment  $\gamma = 0.1$ is chosen. Standard deviation in parentheses. 
\end{table}

\clearpage

\section{A Real-Data Example for Simultaneous Inference in High Dimensions - The Gender Wage Gap Analysis} \label{example}

The following section demonstrates the methods for valid simultaneous inference implemented in the package \Rpackage{hdm} and provides a  comparison of the classical correction methods in a replicable real-data example. 
The gender wage gap, i.e. the relative difference in wages emerging between male and female employees, is a central topic in labor economics. Frequently, studies report an \textit{average} gender wage gap estimate, i.e. how much less women earn as compared to men in terms of average (i.e. mean or median) wages. However, it might be helpful for policy makers to gain a more detailed impression on the gender wage gap and to assess whether and to what extent the gender wage gap differs across individuals. 
A simplistic although frequently encountered approach to assess the wage gap heterogeneity is to compare the relative wage gap across female and male employees in subgroups that are defined in terms of a particular characteristic, e.g. industry. It is obvious that this approach neglects the role of other characteristics relevant for the wage income and, hence, the wage gap, e.g. educational background, experience etc. 
As an examplary illustration, the gender gap in average (mean) earnings in 12 different industrial categories is presented in Figure \ref{indgap}, suggesting that the wage gap differs by large across the subgroups. 
 In contrast to the approach that is simply based on descriptive statistics, an extended regression equation including interaction terms of the gender indicator with observable characteristics is able to take the role of other labor market characteristics into account and, hence, allows to give insights on the determinants of the gender wage gap. As the regression approach leads to a large number of coefficients that are tested simultaneously, an appropriate multiple testing adjustment is required. Thus, the heterogeneous gender wage gap example is used to demonstrate the adjustment methods provided in the \R package \Rpackage{hdm}. The presented example is an illustration of the more extensive analysis of a heterogeneous gender wage gap in \cite{gendergap}.  

\subsection{Data Preparation}


The examplary data is a subsample of the 2016 American Community Survey.\footnote{It can be replicated with the documentation ``Appendix: Replicable Data Example'' that is available online. Further information and the code is provided at \url{https://www.bwl.uni-hamburg.de/en/statistik/forschung/software-und-daten.html}.} 
 The data provide information on civilian full-time working (35+ hours a week, 50+ weeks a year) White, non-Hispanic employees aged older than 25 and younger than 40 with earnings exceeding a the federal minimum level of earnings (\$12,687.5 of yearly wage income).  First, the data set is loaded from the \Rpackage{hdm} package and prepared for the analysis. 
\begin{knitrout}\footnotesize
\definecolor{shadecolor}{rgb}{1, 1, 1}\color{fgcolor}\begin{kframe}
\begin{alltt}
\hlcom{# Load the hdm package}
\hlkwd{rm}\hlstd{(}\hlkwc{list}\hlstd{=}\hlkwd{ls}\hlstd{())}
\hlkwd{library}\hlstd{(hdm)}

\hlcom{# load the ACS data }
\hlkwd{load}\hlstd{(}\hlstr{"ACS2016_gender.rda"}\hlstd{)}
\end{alltt}
\end{kframe}
\end{knitrout}

\begin{knitrout}\footnotesize
\definecolor{shadecolor}{rgb}{1, 1, 1}\color{fgcolor}\begin{figure}

{\centering \includegraphics[width=\linewidth]{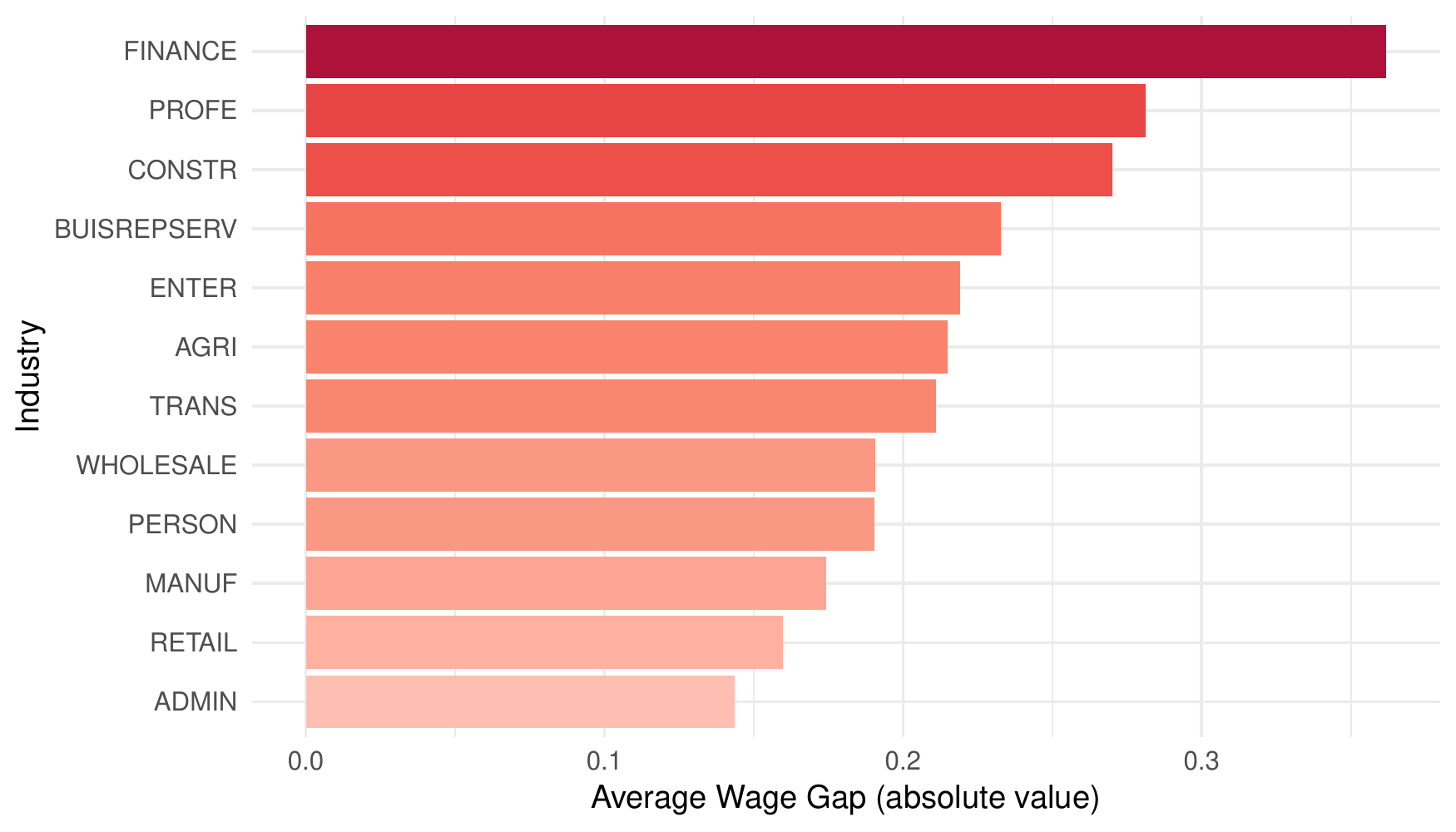} 

}

\caption[\label{indgap} Average Wage Gap across Industries, ACS 2016]{\label{indgap} Average Wage Gap across Industries, ACS 2016.}\label{fig:unnamed-chunk-6}
\end{figure}

\end{knitrout}

\subsection{Valid Simultaneous Inference on a Heterogeneous Gender Wage Gap}
In order to answer the question whether the gender wage gap differs according to the observable characteristics of female employees in a valid way, it is necessary to account for regressors that affect women's job market environment. In the example, variables on marriage, the presence of own children, geographic variation, job characteristics (industry, occupation, hours worked), human capital variables (years of education, experience (squared)), and field of degree are considered. A wage regression is set up that includes all two-way interactions of female with the available characteristics in addition to the baseline regressors, $x_i$.
\begin{align} \label{interact}
\ln{w_i} = \beta_0 + \sum_{k=1}^{K} \theta_k \left(\text{female}_i \times x_{k,i} \right) + x_i'\beta + \varepsilon_i , \quad \quad i = 1,\ldots, n,
\end{align}
The analysis begins with constructing a model matrix that implements the regression relationship of interest. 
\begin{knitrout}\footnotesize
\definecolor{shadecolor}{rgb}{1, 1, 1}\color{fgcolor}\begin{kframe}
\begin{alltt}
\hlcom{# Weekly log wages as outcome variable}
\hlstd{y} \hlkwb{=} \hlstd{data}\hlopt{$}\hlstd{lwwage}

\hlcom{# Model Matrix containing 2-way interaction of female }
\hlcom{# with relevant regressors + covariates}
\hlstd{X} \hlkwb{=} \hlkwd{model.matrix}\hlstd{(} \hlopt{~} \hlnum{1} \hlopt{+} \hlstd{fem} \hlopt{+} \hlstd{fem}\hlopt{:}\hlstd{(ind} \hlopt{+} \hlstd{occ} \hlopt{+} \hlstd{hw} \hlopt{+} \hlstd{deg} \hlopt{+} \hlstd{yos} \hlopt{+} \hlstd{exp} \hlopt{+} \hlstd{exp2} \hlopt{+}
                               \hlstd{married} \hlopt{+} \hlstd{chld19} \hlopt{+} \hlstd{region} \hlopt{+} \hlstd{msa )} \hlopt{+}
                        \hlstd{(married} \hlopt{+}  \hlstd{chld19} \hlopt{+} \hlstd{region} \hlopt{+} \hlstd{msa} \hlopt{+} \hlstd{ind} \hlopt{+} \hlstd{occ} \hlopt{+} \hlstd{hw} \hlopt{+}
                         \hlstd{deg} \hlopt{+} \hlstd{yos} \hlopt{+} \hlstd{exp} \hlopt{+} \hlstd{exp2),} \hlkwc{data} \hlstd{= data)}
\hlcom{# Exclude the constant variables }
\hlstd{X} \hlkwb{=}  \hlstd{X[,}\hlkwd{which}\hlstd{(}\hlkwd{apply}\hlstd{(X,} \hlnum{2}\hlstd{, var)}\hlopt{!=}\hlnum{0}\hlstd{)]}
\hlkwd{dim}\hlstd{(X)}
\end{alltt}
\begin{verbatim}
## [1] 70473   123
\end{verbatim}
\end{kframe}
\end{knitrout}

Accordingly, the regression model considered has $p=123$ regressors in total and is estimated on the basis of $n = 70473$ observations. 
The wage Equation \ref{interact} is estimated with the lasso with the theory-based choice of the penalty term (``rlasso''). To answer the question whether the included regressors have any explanatory power for the outcome variable, the global  test of overall significance is run by calling \texttt{summary()} on the output object of the \texttt{rlasso()} function.  
\begin{knitrout}\footnotesize
\definecolor{shadecolor}{rgb}{1, 1, 1}\color{fgcolor}\begin{kframe}
\begin{alltt}
\hlcom{# Estimate rlasso}
\hlstd{lasso1} \hlkwb{=} \hlkwd{rlasso}\hlstd{(X,y)}

\hlcom{# Run test}
\hlkwd{summary}\hlstd{(lasso1,} \hlkwc{all} \hlstd{=} \hlnum{FALSE}\hlstd{)}

\hlcom{# Output shifted to the Appendix.}
\end{alltt}
\end{kframe}
\end{knitrout}
The hypotheses that all coefficients in the model are zero can be rejected at all common significance levels.  
The main objective of the analysis is to estimate the magnitude of the effects associated with the gender interactions and to assess whether these effects are jointly significantly different from zero.
The so-called ``target'' variables, in total $62$ regressors, are specified in the \texttt{index} option of the \texttt{rlassoEffects()} function. Hence, it is necessary to indicate the columns of the created model matrix that correspond to interactions with the female dummy.

\begin{knitrout}\footnotesize
\definecolor{shadecolor}{rgb}{1, 1, 1}\color{fgcolor}\begin{kframe}
\begin{alltt}
\hlcom{# Construct index for gender variable and interactions  (target parameters)}
\hlstd{index.female} \hlkwb{=} \hlkwd{grep}\hlstd{(}\hlstr{"fem"}\hlstd{,} \hlkwd{colnames}\hlstd{(X))}
\hlstd{K} \hlkwb{=} \hlkwd{length}\hlstd{(index.female)}

\hlcom{# Perform inference on target coefficients}
\hlcom{# estmation might take some time (10 minutes) }
\hlstd{effects} \hlkwb{=} \hlkwd{rlassoEffects}\hlstd{(}\hlkwc{x}\hlstd{=X,}\hlkwc{y}\hlstd{=y,} \hlkwc{index} \hlstd{= index.female)}
\hlkwd{summary}\hlstd{(effects)}

\hlcom{# Output shifted to the Appendix.}
\end{alltt}
\end{kframe}
\end{knitrout}
The output presented in the appendix shows the $K =62$ estimated coefficients together with  t-statistics and unadjusted p-values. 
The next step is to adjust the p-values for multiple testing. Starting with Version \texttt{0.3.0}, the \Rpackage{hdm} offers the S3 method \texttt{p\_adjust()} for objects inheriting from classes \texttt{rlassoEffects} and \texttt{lm}. It hosts the correction methods from the function \texttt{p.adjust()} of the \Rpackage{stats} Package, e.g. Bonferroni, Bonferroni-Holm, Benjamini-Hochberg as well as no correction at all. First,  the naive approach without a multiple testing correction given a  significance level $\alpha$ is presented. Table \ref{rejections} shows the number of rejections at significance levels $\alpha \in \{0.01, 0.05, 0.1 \}$.
\begin{knitrout}\footnotesize
\definecolor{shadecolor}{rgb}{1, 1, 1}\color{fgcolor}\begin{kframe}
\begin{alltt}
\hlcom{# Extract (unadjusted) p-values}
\hlstd{pvals.unadj} \hlkwb{=} \hlkwd{p_adjust}\hlstd{(effects,} \hlkwc{method} \hlstd{=} \hlstr{"none"}\hlstd{)}

\hlcom{# Coefficients and Pvals}
\hlkwd{head}\hlstd{(pvals.unadj)}
\end{alltt}
\begin{verbatim}
##                   Estimate.    pval
## femTRUE             -0.0799 0.20800
## femTRUE:indAGRI     -0.1307 0.00349
## femTRUE:indCONSTR   -0.0521 0.15226
## femTRUE:indMANUF    -0.0097 0.70716
## femTRUE:indTRANS    -0.0403 0.15392
## femTRUE:indRETAIL    0.0264 0.30256
\end{verbatim}
\end{kframe}
\end{knitrout}

Thus, without a correction for multiple testing, $18$, $21$, and $25$ hypotheses could be rejected given the significance levels of 1\%, 5\% and 10\%. 
If one returns to the initial example on variation by industry, one would find significant variation of the wage gap by industry  (as compared to the baseline category ``Wholesale'') in $3$ categories, namely ``Agriculture'', ``Finance, Insurance, and Real Estate'' and ``Professional and Related Services'' at a significance level of 0.1.

Second, classical correction methods like the Bonferroni, Bonferroni-Holm, and the Benjamini-Hochberg adjustments are used to account for testing the $62$ hypotheses at the same time. 

\begin{table}[t]
\begin{tabular}{l r r r } \hline \hline 
Method & \multicolumn{3}{c}{Significance Level} \\ 
& 0.01 & 0.05 & 0.10 \\[0.8ex] \hline
    Naive &  $18$ &  $21$ &  $25$ \\ 
    Benjamini-Hochberg & $10$ &  $19$ &  $21$ \\ 
    Bonferroni & $9$ &  $10$ &  $10$ \\ 
    Holm & $9$ &  $10$ &  $10$ \\ 
    Joint Confidence Region & $9$ &  $10$ &  $11$ \\
    Romano-Wolf & $8$ &  $11$ &  $11$     \\[0.8ex] \hline   \hline
\end{tabular}
\caption{Number of Rejected Hypotheses}
\label{rejections}
\end{table}

\begin{knitrout}\footnotesize
\definecolor{shadecolor}{rgb}{1, 1, 1}\color{fgcolor}\begin{kframe}
\begin{alltt}
\hlcom{# Bonferroni }
\hlstd{pvals.bonf} \hlkwb{=} \hlkwd{p_adjust}\hlstd{(effects,} \hlkwc{method} \hlstd{=} \hlstr{"bonferroni"}\hlstd{)}

\hlcom{# Holm}
\hlstd{pvals.holm} \hlkwb{=} \hlkwd{p_adjust}\hlstd{(effects,} \hlkwc{method} \hlstd{=} \hlstr{"holm"}\hlstd{)}

\hlkwd{head}\hlstd{(pvals.bonf)}
\end{alltt}
\begin{verbatim}
##                   Estimate.  pval
## femTRUE             -0.0799 1.000
## femTRUE:indAGRI     -0.1307 0.217
## femTRUE:indCONSTR   -0.0521 1.000
## femTRUE:indMANUF    -0.0097 1.000
## femTRUE:indTRANS    -0.0403 1.000
## femTRUE:indRETAIL    0.0264 1.000
\end{verbatim}
\begin{alltt}
\hlkwd{head}\hlstd{(pvals.holm)}
\end{alltt}
\begin{verbatim}
##                   Estimate.  pval
## femTRUE             -0.0799 1.000
## femTRUE:indAGRI     -0.1307 0.178
## femTRUE:indCONSTR   -0.0521 1.000
## femTRUE:indMANUF    -0.0097 1.000
## femTRUE:indTRANS    -0.0403 1.000
## femTRUE:indRETAIL    0.0264 1.000
\end{verbatim}
\end{kframe}
\end{knitrout}
As a general improvement, the Holm-corrected p-values are smaller or equal to those obtained from a Bonferroni adjustment. At significance levels 1\%, 5\% and 10\%, it is possible to reject fewer hypotheses if p-values are corrected for multiple testing.

According to the Benjamini-Hochberg (BH) correction  \citep{bh1995} of p-values that achieves control of the FDR it is possible to reject $10, 19,$ and $21$ null hypotheses at specified values of the FDR, $\gamma$, at 0.01, 0.05 and 0.1. 
\begin{knitrout}\footnotesize
\definecolor{shadecolor}{rgb}{1, 1, 1}\color{fgcolor}\begin{kframe}
\begin{alltt}
\hlstd{pvals.BH} \hlkwb{=} \hlkwd{p_adjust}\hlstd{(effects,} \hlkwc{method} \hlstd{=} \hlstr{"BH"}\hlstd{)}
\hlkwd{head}\hlstd{(pvals.BH)}
\end{alltt}
\begin{verbatim}
##                   Estimate.   pval
## femTRUE             -0.0799 0.3778
## femTRUE:indAGRI     -0.1307 0.0174
## femTRUE:indCONSTR   -0.0521 0.3078
## femTRUE:indMANUF    -0.0097 0.8412
## femTRUE:indTRANS    -0.0403 0.3078
## femTRUE:indRETAIL    0.0264 0.4690
\end{verbatim}
\end{kframe}
\end{knitrout}

Regarding variation by industry, the Bonferroni and Holm procedure find a significantly different wage gap (at the 10\% significance level) only for industry ``Finance, Insurance, and Real Estate'' whereas the Benjamini-Hochberg correction with $\gamma = 0.1$ leads to the same conclusion as obtained without any adjustment. 
These results can now be compared to results obtained from joint significance test with and without the Romano-Wolf stepdown procedure. We can start with construction of a joint $0.9$-confidence region for the $62$ coefficients using the option \texttt{joint=T} in the \texttt{confint()} function for objects of the class \texttt{rlassoEffects}. 
\begin{knitrout}\footnotesize
\definecolor{shadecolor}{rgb}{1, 1, 1}\color{fgcolor}\begin{kframe}
\begin{alltt}
\hlcom{# valid joint 0.95-confidence interval}
\hlstd{alpha} \hlkwb{=} \hlnum{0.1}

\hlstd{CI} \hlkwb{=} \hlkwd{confint}\hlstd{(effects,} \hlkwc{level} \hlstd{=} \hlnum{1}\hlopt{-}\hlstd{alpha,} \hlkwc{joint}\hlstd{=T)}
\hlkwd{head}\hlstd{(CI)}
\end{alltt}
\begin{verbatim}
##                       5 %   95 %
## femTRUE           -0.2758 0.1160
## femTRUE:indAGRI   -0.2923 0.0310
## femTRUE:indCONSTR -0.1632 0.0589
## femTRUE:indMANUF  -0.0912 0.0718
## femTRUE:indTRANS  -0.1305 0.0499
## femTRUE:indRETAIL -0.0582 0.1110
\end{verbatim}
\end{kframe}
\end{knitrout}
The results from the confidence intervals are equivalent to a test at significance level  $\alpha=0.1$ so that $11$ hypotheses can be rejected. 
However, the Romano-Wolf stepdown procedure allows to increase power. For instance with a significance level of 5\%, the stepdown correction allows to reject one hypothesis more than with the joint confidence interval. The p-values can be adjusted according to the Romano-Wolf-stepdown algorithm by setting the option \texttt{method = }``RW'' (default) of the \texttt{p\_adjust()} call. The number of repetitions can be varied by specifying the option \texttt{B}, $B=1000$ by default. 
\begin{knitrout}\footnotesize
\definecolor{shadecolor}{rgb}{1, 1, 1}\color{fgcolor}\begin{kframe}
\begin{alltt}
\hlcom{# Romano-Wolf stepdown adjustment}
\hlstd{pvals.RW} \hlkwb{=} \hlkwd{p_adjust}\hlstd{(effects,} \hlkwc{method} \hlstd{=} \hlstr{"RW"}\hlstd{,} \hlkwc{B}\hlstd{=}\hlnum{1000}\hlstd{)}

\hlkwd{head}\hlstd{(pvals.RW)}
\end{alltt}
\begin{verbatim}
##                   Estimate.  pval
## femTRUE             -0.0799 0.997
## femTRUE:indAGRI     -0.1307 0.369
## femTRUE:indCONSTR   -0.0521 0.986
## femTRUE:indMANUF    -0.0097 1.000
## femTRUE:indTRANS    -0.0403 0.992
## femTRUE:indRETAIL    0.0264 0.999
\end{verbatim}
\end{kframe}
\end{knitrout}
Using the joint confidence interval and the Romano-Wolf stepdown adjustment allows to reject more hypotheses than with traditional methods at significance levels 5\% and 10\%. Hence, taking into account the dependence of test statistics is beneficial in terms of power in the real-data example. 

\section{Conclusion}
The previous sections provide a short overview on important methods for multiple testing adjustment in a high-dimensional regression setting. Throughout the paper, our intention was to present the concepts and the necessity of a multiple adjustment in a comprehensive way. Similarly, the tools for valid simultaneous inference in high-dimensional settings that are available in the \R package \Rpackage{hdm} are intended to be easy to use in empirical applications. The demonstration of the methods in the real-data example are intended to motivate applied statisticians to (i) use modern statistical methods for high-dimensional regression, i.e. the lasso, and (ii) to appropriately adjust if multiple hypotheses are tested simultaneously. Since the \Rpackage{hdm} provides user-friendly adjustment methods for objects of the S3 class \texttt{lm}, we hope that uses will use the correction methods more frequently even in classical least squares regression.

\clearpage

\section{Appendix}

\subsection{Details of Simulation Study}

The simulation study was implemented using the statistical software \R (R version 3.3.3 (2017-03-06)) on a x86\_64-redhat-linux-gnu (64 bit) platform. For the sake of replicability, the R code for the simulation study is available as supplemental material on the website \url{https://www.bwl.uni-hamburg.de/en/statistik/forschung/software-und-daten.html}. In the lasso regression, the theory-based and data-dependent choice of the penalty term $\lambda$ for homoscedastic errors is implemented \citep{rjournalhdm}.

\subsection{Additional Simulation Results} 

We additionally provide the simulation results if significance tests are based on a level of $\alpha = 0.05$ or $\alpha = 0.01$.

\clearpage

\begin{table}[t]
\begin{tabular}{l r r r r r } \hline \hline 
   & Naive & Benjamini-Hochberg & Bonferroni & Bonferroni-Holm & Romano-Wolf \\[0.8ex]
     \hline   \hline 
   \multicolumn{6}{c}{Correct Rejections} \\[0.8ex] \hline 
 $n = 200$ & $10.201$ & $8.730$ & $7.163$ & $7.231$ & $7.293$ \\
  & $(0.842)$ & $(1.008)$ & $(0.974)$ & $(0.981)$ & $(0.983)$  \\
  & & & & & \\
  \hline 
\multicolumn{6}{c}{Incorrect Rejections}  \\ \hline 
    & $2.676$ & $0.545$ & $0.071$ & $0.078$ & $0.086$ \\
  & $(1.758)$ & $(0.836)$ & $(0.275)$ & $(0.290)$ & $(0.305)$ \\
  & & & & & \\
  \hline 
  \multicolumn{6}{c}{Familywise Error Rate}  \\ \hline 
     & $0.916$ & $0.381$ & $0.067$ & $0.073$ & $0.080$ \\
  & & & & & \\
  \hline 
  \multicolumn{6}{c}{False Discovery Rate}  \\ \hline 
     & $0.194$ & $0.052$ & $0.009$ & $0.009$ & $0.010$ \\
  & & & & & \\ \hline   \hline
  \multicolumn{6}{c}{Correct Rejections} \\[0.8ex] \hline
 $n = 500$ & $11.770$ & $11.350$ & $10.364$ & $10.441$ & $10.483$ \\
  & $(0.440)$ & $(0.648)$ & $(0.781)$ & $(0.775)$ & $(0.771)$  \\
  & & & & & \\
  \hline
\multicolumn{6}{c}{Incorrect Rejections}  \\ \hline
    & $2.518$ & $0.588$ & $0.053$ & $0.062$ & $0.068$ \\
  & $(1.721)$ & $(0.888)$ & $(0.240)$ & $(0.260)$ & $(0.272)$ \\
  & & & & & \\
  \hline 
  \multicolumn{6}{c}{Familywise Error Rate}  \\ \hline
     & $0.898$ & $0.394$ & $0.050$ & $0.058$ & $0.062$ \\
    & & & & & \\
  \hline 
  \multicolumn{6}{c}{False Discovery Rate}  \\ \hline 
    & $0.165$ & $0.045$ & $0.005$ & $0.005$ & $0.006$  \\[0.8ex] \hline   \hline
\end{tabular}
\caption{Simulation Results}
\label{simresults2}
The Table presents the average number of correct or incorrect rejections at a significance level $\alpha = 0.05$ and the FWER over all $R = 5000$ repetitions. For the  Benjamini-Hochberg adjustment  $\gamma = 0.05$ is chosen. Standard deviation in parantheses. 
\end{table}

\begin{table}[t]
\begin{tabular}{l r r r r r } \hline \hline 
   & Naive & Benjamini-Hochberg & Bonferroni & Bonferroni-Holm & Romano-Wolf \\[0.8ex]
     \hline   \hline 
   \multicolumn{6}{c}{Correct Rejections} \\[0.8ex] \hline 
 $n = 200$ & $8.905$ & $7.446$ & $6.294$ & $6.346$ & $6.390$ \\
  & $(0.930)$ & $(1.041)$ & $(0.940)$ & $(0.949)$ & $(0.961)$  \\
  & & & & & \\
  \hline 
\multicolumn{6}{c}{Incorrect Rejections}  \\ \hline 
    & $0.614$ & $0.113$ & $0.019$ & $0.022$ & $0.023$ \\
  & $(0.841)$ & $(0.355)$ & $(0.140)$ & $(0.152)$ & $(0.155)$ \\
  & & & & & \\
  \hline 
  \multicolumn{6}{c}{Familywise Error Rate}  \\ \hline 
     & $0.434$ & $0.101$ & $0.019$ & $0.021$ & $0.022$ \\
 & & & & & \\
  \hline 
  \multicolumn{6}{c}{False Discovery Rate}  \\ \hline 
     & $0.059$ & $0.013$ & $0.003$ & $0.003$ & $0.003$ \\
    & & & & & \\ \hline   \hline
  \multicolumn{6}{c}{Correct Rejections} \\[0.8ex] \hline
 $n = 500$ & $11.342$ & $10.706$ & $9.642$ & $9.716$ & $9.746$ \\
  & $(0.641)$ & $(0.763)$ & $(0.803)$ & $(0.811)$ & $(0.828)$  \\
  & & & & & \\
  \hline
\multicolumn{6}{c}{Incorrect Rejections}  \\ \hline
    & $0.537$ & $0.119$ & $0.012$ & $0.014$ & $0.015$ \\
  & $(0.792)$ & $(0.362)$ & $(0.110)$ & $(0.119)$ & $(0.126)$ \\
  & & & & & \\
  \hline 
  \multicolumn{6}{c}{Familywise Error Rate}  \\ \hline
     & $0.389$ & $0.107$ & $0.012$ & $0.014$ & $0.015$ \\
  & & & & & \\
  \hline 
  \multicolumn{6}{c}{False Discovery Rate}  \\ \hline 
    & $0.042$ & $0.010$ & $0.001$ & $0.001$ & $0.001$  \\[0.8ex] \hline   \hline
\end{tabular}
\caption{Simulation Results}
\label{simresults3}
The Table presents the average number of correct or incorrect rejections at a significance level $\alpha = 0.01$ and the FWER over all $R = 5000$ repetitions. For the  Benjamini-Hochberg adjustment  $\gamma = 0.01$ is chosen. Standard deviation in parantheses. 
\end{table}

\clearpage
\subsection{Additional Results, Real-Data Example}
For the sake of brevity, the full summary outputs of the global significance test and the joint significance tes for the target coefficients with lasso and double selection are omitted in the main text. For completeness, the output is presented in the following.

\begin{knitrout}\tiny
\definecolor{shadecolor}{rgb}{1, 1, 1}\color{fgcolor}\begin{kframe}
\begin{alltt}
\hlstd{lasso1} \hlkwb{=} \hlkwd{rlasso}\hlstd{(X,y)}

\hlcom{# Run test}
\hlkwd{summary}\hlstd{(lasso1,} \hlkwc{all} \hlstd{=} \hlnum{FALSE}\hlstd{)}
\end{alltt}
\begin{verbatim}
## 
## Call:
## rlasso.default(x = X, y = y)
## 
## Post-Lasso Estimation:  TRUE 
## 
## Total number of variables: 123
## Number of selected variables: 58 
## 
## Residuals: 
##      Min       1Q   Median       3Q      Max 
## -2.50948 -0.27489 -0.00787  0.25579  2.66745 
## 
##                                    Estimate
## (Intercept)                            4.70
## femTRUE                                0.01
## married                                0.12
## chld19                                 0.09
## regionMiddle Atlantic Division         0.07
## regionEast North Central Div.         -0.04
## regionWest North Central Div.         -0.07
## regionEast South Central Div.         -0.12
## regionMountain Division               -0.06
## regionPacific Division                 0.13
## msa                                    0.18
## indAGRI                               -0.22
## indMANUF                               0.08
## indTRANS                               0.09
## indRETAIL                             -0.14
## indFINANCE                             0.18
## indBUISREPSERV                         0.12
## indENTER                              -0.08
## indPROFE                              -0.06
## indADMIN                              -0.05
## occBus Operat Spec                    -0.04
## occComput/Math                         0.02
## occLife/Physical/Soc Sci.             -0.18
## occComm/Soc Serv                      -0.34
## occLegal                               0.11
## occEduc/Training/Libr                 -0.35
## occArts/Design/Entert/Sports/Media    -0.17
## occHealthc Pract/Technic               0.10
## occProtect Serv                       -0.07
## occOffice/Administr Supp              -0.32
## occProd                               -0.32
## hw50to59                               0.21
## hw60to69                               0.28
## hw70plus                               0.25
## degComp/Inform Sci                     0.16
## degEngin                               0.21
## degEnglish/Lit/Compos                 -0.06
## degLib Arts/Hum                       -0.06
## degBio/Life Sci                        0.04
## degMath/Stats                          0.16
## degPhys Fit/Parks/Recr/Leis           -0.08
## degPsych                              -0.04
## degCrim Just/Fire Prot                -0.04
## degPubl Aff/Policy/Soc Wo             -0.05
## degSoc Sci                             0.08
## degFine Arts                          -0.08
## degBus                                 0.08
## degHist                               -0.06
## yos                                    0.11
## exp                                    0.03
## femTRUE:married                       -0.05
## femTRUE:chld19                        -0.04
## femTRUE:regionMountain Division       -0.01
## femTRUE:indAGRI                       -0.07
## femTRUE:indFINANCE                    -0.13
## femTRUE:occArchit/Engin                0.03
## femTRUE:occOffice/Administr Supp      -0.02
## femTRUE:degPubl Aff/Policy/Soc Wo     -0.01
## femTRUE:exp2                          -0.02
## 
## Residual standard error: 0.463
## Multiple R-squared:  0.391
## Adjusted R-squared:  0.39
## Joint significance test:
##  the sup score statistic for joint significance test is  280 with a p-value of    0
\end{verbatim}
\end{kframe}
\end{knitrout}

\begin{knitrout}\tiny
\definecolor{shadecolor}{rgb}{1, 1, 1}\color{fgcolor}\begin{kframe}
\begin{alltt}
\hlcom{# Summary of significance test }
\hlkwd{summary}\hlstd{(effects)}
\end{alltt}
\begin{verbatim}
## [1] "Estimates and significance testing of the effect of target variables"
##                                            Estimate. Std. Error t value
## femTRUE                                    -0.079926   0.063479   -1.26
## femTRUE:indAGRI                            -0.130650   0.044735   -2.92
## femTRUE:indCONSTR                          -0.052142   0.036422   -1.43
## femTRUE:indMANUF                           -0.009699   0.025818   -0.38
## femTRUE:indTRANS                           -0.040300   0.028265   -1.43
## femTRUE:indRETAIL                           0.026415   0.025622    1.03
## femTRUE:indFINANCE                         -0.135372   0.024672   -5.49
## femTRUE:indBUISREPSERV                     -0.035614   0.026010   -1.37
## femTRUE:indPERSON                          -0.047695   0.039551   -1.21
## femTRUE:indENTER                           -0.050513   0.039630   -1.27
## femTRUE:indPROFE                           -0.054828   0.024235   -2.26
## femTRUE:indADMIN                           -0.011253   0.028358   -0.40
## femTRUE:occBus Operat Spec                  0.025938   0.016180    1.60
## femTRUE:occFinanc Spec                     -0.048161   0.016859   -2.86
## femTRUE:occComput/Math                     -0.006376   0.018829   -0.34
## femTRUE:occArchit/Engin                     0.043310   0.027374    1.58
## femTRUE:occLife/Physical/Soc Sci.           0.061113   0.024642    2.48
## femTRUE:occComm/Soc Serv                    0.156487   0.024372    6.42
## femTRUE:occLegal                            0.009461   0.021722    0.44
## femTRUE:occEduc/Training/Libr               0.115386   0.015943    7.24
## femTRUE:occArts/Design/Entert/Sports/Media  0.046201   0.019747    2.34
## femTRUE:occHealthc Pract/Technic            0.006586   0.017898    0.37
## femTRUE:occProtect Serv                    -0.003907   0.036230   -0.11
## femTRUE:occSales                           -0.018405   0.015144   -1.22
## femTRUE:occOffice/Administr Supp           -0.000996   0.015553   -0.06
## femTRUE:occProd                             0.001608   0.036690    0.04
## femTRUE:hw40to49                           -0.053407   0.017716   -3.01
## femTRUE:hw50to59                           -0.071987   0.019040   -3.78
## femTRUE:hw60to69                           -0.123902   0.022724   -5.45
## femTRUE:hw70plus                           -0.202611   0.031436   -6.45
## femTRUE:degAgri                             0.008232   0.035918    0.23
## femTRUE:degComm                             0.039552   0.021445    1.84
## femTRUE:degComp/Inform Sci                 -0.080594   0.030006   -2.69
## femTRUE:degEngin                           -0.007807   0.025530   -0.31
## femTRUE:degEnglish/Lit/Compos               0.019353   0.024121    0.80
## femTRUE:degLib Arts/Hum                     0.046775   0.037581    1.24
## femTRUE:degBio/Life Sci                    -0.037057   0.022191   -1.67
## femTRUE:degMath/Stats                      -0.063706   0.034123   -1.87
## femTRUE:degPhys Fit/Parks/Recr/Leis        -0.022403   0.030424   -0.74
## femTRUE:degPhys Sci                        -0.075693   0.026715   -2.83
## femTRUE:degPsych                           -0.007874   0.023112   -0.34
## femTRUE:degCrim Just/Fire Prot             -0.085437   0.029390   -2.91
## femTRUE:degPubl Aff/Policy/Soc Wo          -0.017049   0.041864   -0.41
## femTRUE:degSoc Sci                         -0.052175   0.019783   -2.64
## femTRUE:degFine Arts                       -0.017200   0.022201   -0.77
## femTRUE:degMed/Hlth Sci Serv               -0.026064   0.025231   -1.03
## femTRUE:degBus                             -0.020136   0.017776   -1.13
## femTRUE:degHist                            -0.071948   0.026570   -2.71
## femTRUE:yos                                 0.005787   0.003359    1.72
## femTRUE:exp                                -0.001838   0.003762   -0.49
## femTRUE:exp2                               -0.008556   0.009151   -0.93
## femTRUE:married                            -0.050840   0.009097   -5.59
## femTRUE:chld19                             -0.051834   0.009268   -5.59
## femTRUE:regionMiddle Atlantic Division     -0.024007   0.015626   -1.54
## femTRUE:regionEast North Central Div.      -0.009862   0.015658   -0.63
## femTRUE:regionWest North Central Div.      -0.011769   0.018628   -0.63
## femTRUE:regionSouth Atlantic Division      -0.001011   0.015414   -0.07
## femTRUE:regionEast South Central Div.       0.006562   0.020696    0.32
## femTRUE:regionWest South Central Div.      -0.072252   0.017598   -4.11
## femTRUE:regionMountain Division            -0.028934   0.019022   -1.52
## femTRUE:regionPacific Division             -0.057981   0.016211   -3.58
## femTRUE:msa                                -0.002253   0.014194   -0.16
##                                            Pr(>|t|)    
## femTRUE                                     0.20800    
## femTRUE:indAGRI                             0.00349 ** 
## femTRUE:indCONSTR                           0.15226    
## femTRUE:indMANUF                            0.70716    
## femTRUE:indTRANS                            0.15392    
## femTRUE:indRETAIL                           0.30256    
## femTRUE:indFINANCE                          4.1e-08 ***
## femTRUE:indBUISREPSERV                      0.17091    
## femTRUE:indPERSON                           0.22785    
## femTRUE:indENTER                            0.20245    
## femTRUE:indPROFE                            0.02368 *  
## femTRUE:indADMIN                            0.69150    
## femTRUE:occBus Operat Spec                  0.10892    
## femTRUE:occFinanc Spec                      0.00428 ** 
## femTRUE:occComput/Math                      0.73491    
## femTRUE:occArchit/Engin                     0.11361    
## femTRUE:occLife/Physical/Soc Sci.           0.01314 *  
## femTRUE:occComm/Soc Serv                    1.4e-10 ***
## femTRUE:occLegal                            0.66316    
## femTRUE:occEduc/Training/Libr               4.6e-13 ***
## femTRUE:occArts/Design/Entert/Sports/Media  0.01930 *  
## femTRUE:occHealthc Pract/Technic            0.71291    
## femTRUE:occProtect Serv                     0.91413    
## femTRUE:occSales                            0.22424    
## femTRUE:occOffice/Administr Supp            0.94896    
## femTRUE:occProd                             0.96504    
## femTRUE:hw40to49                            0.00257 ** 
## femTRUE:hw50to59                            0.00016 ***
## femTRUE:hw60to69                            5.0e-08 ***
## femTRUE:hw70plus                            1.2e-10 ***
## femTRUE:degAgri                             0.81872    
## femTRUE:degComm                             0.06514 .  
## femTRUE:degComp/Inform Sci                  0.00723 ** 
## femTRUE:degEngin                            0.75976    
## femTRUE:degEnglish/Lit/Compos               0.42237    
## femTRUE:degLib Arts/Hum                     0.21327    
## femTRUE:degBio/Life Sci                     0.09494 .  
## femTRUE:degMath/Stats                       0.06191 .  
## femTRUE:degPhys Fit/Parks/Recr/Leis         0.46152    
## femTRUE:degPhys Sci                         0.00461 ** 
## femTRUE:degPsych                            0.73334    
## femTRUE:degCrim Just/Fire Prot              0.00365 ** 
## femTRUE:degPubl Aff/Policy/Soc Wo           0.68383    
## femTRUE:degSoc Sci                          0.00836 ** 
## femTRUE:degFine Arts                        0.43848    
## femTRUE:degMed/Hlth Sci Serv                0.30160    
## femTRUE:degBus                              0.25730    
## femTRUE:degHist                             0.00677 ** 
## femTRUE:yos                                 0.08494 .  
## femTRUE:exp                                 0.62519    
## femTRUE:exp2                                0.34980    
## femTRUE:married                             2.3e-08 ***
## femTRUE:chld19                              2.2e-08 ***
## femTRUE:regionMiddle Atlantic Division      0.12446    
## femTRUE:regionEast North Central Div.       0.52878    
## femTRUE:regionWest North Central Div.       0.52751    
## femTRUE:regionSouth Atlantic Division       0.94770    
## femTRUE:regionEast South Central Div.       0.75119    
## femTRUE:regionWest South Central Div.       4.0e-05 ***
## femTRUE:regionMountain Division             0.12824    
## femTRUE:regionPacific Division              0.00035 ***
## femTRUE:msa                                 0.87386    
## ---
## Signif. codes:  0 '***' 0.001 '**' 0.01 '*' 0.05 '.' 0.1 ' ' 1
\end{verbatim}
\end{kframe}
\end{knitrout}

\clearpage 
\footnotesize
\bibliographystyle{econometrica}
\bibliography{mybib_simultaneousInf}

\end{document}